\documentclass[prd,groupedaddress,twocolumn]{revtex4-1}
\usepackage{graphicx}
\usepackage{graphics}
\usepackage{amsmath}
\usepackage{amssymb}
\usepackage{mathtools} 
\begin{document}

%\preprint{}

\title{ Interparticle potential energy for $D$-dimensional electromagnetic models from the corresponding scalar ones} 


\author{Antonio Accioly}

\email[]{accioly@cbpf.br}

\affiliation{Laborat\'{o}rio de F\'{\i}sica Experimental (LAFEX), Centro Brasileiro de Pesquisas F\'{i}sicas (CBPF), Rua Dr. Xavier Sigaud 150, Urca, 22290-180, Rio de Janeiro, RJ, Brazil}

\author{Jos\'{e} Helay\"{e}l-Neto}
\email[]{helayel@cbpf.br}

\affiliation{Laborat\'{o}rio de F\'{\i}sica Experimental (LAFEX), Centro Brasileiro de Pesquisas F\'{i}sicas (CBPF), Rua Dr. Xavier Sigaud 150, Urca, 22290-180, Rio de Janeiro, RJ, Brazil}

\author{Gilson Correia}

\email[]{gilson@cbpf.br}

\affiliation{Laborat\'{o}rio de F\'{\i}sica Experimental (LAFEX), Centro Brasileiro de Pesquisas F\'{i}sicas (CBPF), Rua Dr. Xavier Sigaud 150, Urca, 22290-180, Rio de Janeiro, RJ, Brazil}

\author{Gustavo Brito}

\email[]{gpbrito@cbpf.br}

\affiliation{Laborat\'{o}rio de F\'{\i}sica Experimental (LAFEX), Centro Brasileiro de Pesquisas F\'{i}sicas (CBPF), Rua Dr. Xavier Sigaud 150, Urca, 22290-180, Rio de Janeiro, RJ, Brazil}

\author{Jos\'{e} de Almeida}

\email[]{josejr@cbpf.br}

\affiliation{Laborat\'{o}rio de F\'{\i}sica Experimental (LAFEX), Centro Brasileiro de Pesquisas F\'{i}sicas (CBPF), Rua Dr. Xavier Sigaud 150, Urca, 22290-180, Rio de Janeiro, RJ, Brazil}

\author{Wallace Herdy }
\email[]{wallacew@cbpf.br}

\affiliation{Laborat\'{o}rio de F\'{\i}sica Experimental (LAFEX), Centro Brasileiro de Pesquisas F\'{i}sicas (CBPF), Rua Dr. Xavier Sigaud 150, Urca, 22290-180, Rio de Janeiro, RJ, Brazil}

%\thanks{}


\date{\today}


\begin{abstract} 
Using a method based on the generating functional plus a  kind of ``correspondence principle"  --- which acts as a bridge between the electromagnetic and scalar fields --- it is shown that the interparticle potential energy concerning  a  given $D$-dimensional electromagnetic model can be obtained in a simple way from that related to the corresponding  scalar system. The  $D$-dimensional electromagnetic potential for a general model containing higher derivatives is then found from the corresponding scalar one and the behavior of the former  is analyzed at large as well as small distances.  In addition, we investigate   the presence of ghosts in   the  four-dimensional version of the potential associated  with the model above  and analyze the reason why  the Coulomb singularity is absent from this system.  The no-go theorem by Ostrogradski is demystified as well.

\end{abstract}
%\keywords{11.10.Kk, 11.15.-q, 12.60.Cn, 14.70.Pw}



\maketitle

\section{Introduction}

Time and again new interesting and important electromagnetic models are proposed aiming to overcome the hurdles that are  inherent in the theoretical description of the electromagnetic interactions. Nonetheless, all these systems have inevitably to reproduce in the nonrelativistic limit the Coulombian potential energy plus a  possible correction. Now, bearing in mind that  this potential energy is singular at the origin, it is easy to understand the great importance of searching for  electromagnetic models in which this singularity is absent. Accordingly, it is extremely important to have available an easy prescription on hand for finding the potential energy for those new electromagnetic models  so that their  behavior at small distances can be analyzed promptly and efficiently.

In this vein, a simple prescription for computing the mentioned potential was recently built up \cite{1}. Our primary aim here is to show that if we start from the scalar theory corresponding to the electromagnetic model we want to analyze and  which is obtained by utilizing a kind of ``corespondence principle" that acts as a bridge between the electromagnetic and scalar fields, we will arrive, {\it mutatis mutandis}, at the potential for the electromagnetic
system described in \cite{1}. Of course, to work with scalar theories is always much easier than with electromagnetic ones, which is a good argument in favor of our method.

On the other hand, electromagnetic theories lacking  the Coulomb singularity can often be obtained  by adding higher-order terms to the Maxwell Lagrangian. Why is this so? Because the higher-derivative terms are responsible for giving origin to a potential with a sign that is opposite to the Coulomb one and, as a result, at the origin   this correction to the Coulomb potential is responsible in  general for canceling  out the contribution coming from the  aforementioned  potential. Now, since the higher-derivative potential contributes with an  energy that has  a sign which is the opposite of  that concerning  the photon, we are in the presence of a ghost. Note, however, that renormalizable higher-order  theories can be seen in general  as very satisfactory effective   field theories below the Planck scale \cite{2,3,4}. Our second aim in this paper is to discuss the issue of  ghosts in higher-derivative   theories. In particular, we shall discuss why the Ostrogradski's no-go theorem  \cite{5} cannot be used to discard  gauge theories.

The article is organized as follows.

 In the next section we discuss  the method for computing the  $D$-dimensional   electromagnetic interparticle potential energy from the corresponding  scalar one.

In Sec. III, we compute the $D$-dimensional interparticle  potential energy for a general  electromagnetic  model containing higher derivatives utilizing the related scalar system and analyze the behavior of the former  both at large and small distances.

%\begin{equation}
%{\cal{L}}= - \frac{1}{4}F^2_{\mu \nu}  + \frac{1}{2}M^2 A^2_{\mu}  - \frac{1}{4m^2}F_{\mu \nu} \Box F^{\mu \nu} - J^\mu A_\mu,
%\end{equation} 

%\noindent where $M$ ($> 0$) and $m$ ($ >0$) are parameters with mass dimension, via the corresponding  scalar theory, i.e., 

%\begin{equation}
%{\cal{L}}= \frac{1}{2}\partial_\mu \phi \partial^\mu \phi - \frac{1}{2}M^2 \phi^2 + \frac{1}{2m^2}\partial_\mu \phi  \Box \partial^\mu \phi  + J\phi.
%\end{equation}

   We  investigate in Sec. IV the issue of ghosts in higher-derivative models and  demystify the no-go theorem by Ostrogradski. In particular, we discuss the  presence of ghosts in   the  four-dimensional version of the potential found in Sec. III and   analyze the reason why the Coulomb singularity is absent from the aforementioned electromagnetic   systems.

Our conclusions are presented in Sec. V.

Technical details are relegated to the Appendix.

We use natural units throughout and our Minkowski metric is diag(1, -1, ..., -1).

\section{ THE METHOD }

As is well known,  the generating functional for the  connected Feynman diagrams $W_D(J)$ is related to the generating functional $Z_D(J)$ for a scalar theory,  by   $Z_D(J)= e^{i W_D(J)}$ \cite{6,7,8},
\noindent where 

\begin{eqnarray}
W_D(J)= -\frac{1}{2}\int \int{d^D x d^D y J(x) D(x-y)J(y) }.
\end{eqnarray}  

\noindent Here $J(x)$ and $D(x-y)$ are, respectively, the external  current and the propagator.

Now, keeping in mind that 

\begin{eqnarray}
D(x-y)&=& \int{\frac{d^D k}{(2\pi)^D}e^{ik(x-y)}D(k)},\nonumber \\ J(k)&=& \int{d^D x e^{-ikx}}J(x), \nonumber
\end{eqnarray}

\noindent we promptly obtain

\begin{eqnarray}
W_D(J)= -\frac{1}{2} \int{\frac{d^D k}{(2\pi)^D} J(k)^*D(k)J(k)}.
\end {eqnarray}

 Assuming then that the external current is time independent, we get from (2)

\begin{eqnarray}
W_D(J)&=&- \frac{1}{2} \int \frac{d^D k}{(2 \pi)^{D-1}} \Big[ \delta (k^0) \; T \;  D(k)  \int \int d^{D-1} {\bf{x}}   
\nonumber \\ &&\times  \; d^{D-1} {\bf{y}}
   e^{i{\bf{k}}\cdot{\bf{(y-x})}}J({ \bf x)}J({\bf y}) \Big],
\end{eqnarray}

\noindent where the time interval $T$ is produced by the factor $\int{dx^0}$.

Simple algebraic  manipulations, on the other hand, reduces (3) to the form

\begin{eqnarray}
W_D(J)= -T\int{\frac{d^{D-1} {\bf{k}}}{(2 \pi)^{D-1}} D({\bf k}) \Delta({\bf{k}})},
\end{eqnarray}

\noindent where  $ D({\bf k}) \equiv  D(k)|_{k^0=0} $,  and

\begin{eqnarray}
\Delta({\bf{k}}) \equiv \int \int d^{D-1} {\bf{x}} d^{D-1} {\bf{y}} e^{i {\bf{k}} \cdot ({\bf{y-x})}} \frac{J( {\bf x}) J({ \bf y})}{2}.
\end{eqnarray}

 In the specific case of two  scalar charges $\sigma_1$ and $\sigma_2$ located, respectively, at ${\bf{a_1}}$ and ${\bf{a_2}}$, the current assumes the form

\begin{eqnarray}
J({\bf x}) =  \sigma_1 \delta^{D-1}({\bf{x - a_1}})
 + \sigma_2 \delta^{D-1} ({\bf{x - a_2}}).
\end{eqnarray}

Therefore, 

\begin{eqnarray}
\Delta({\bf{k}})= \sigma_1 \sigma_2 e^{i {\bf{k}} \cdot {\bf{r}}},
\end{eqnarray}

\noindent where ${\bf{r= a_2 - a_1}}$, and 

\begin{eqnarray}
W_D(J)= - T\frac{\sigma_1 \sigma_2}{(2 \pi)^{D-1}} \int{d^{D-1}{\bf{k}}e^{i {\bf{k}} \cdot {\bf{r}}} D({\bf{k}})}.
\end{eqnarray}

Bearing in mind that

\begin{eqnarray}
Z_D(J)= <0\big|e^{-iH_D T} \big|0> = e^{-iE_D^{(\mathrm{scal})} T},
\end{eqnarray}

\noindent which implies that 

\begin{eqnarray}
E_D^{({\mathrm{scal}})} = - \frac{W_D(J)}{T},
\end{eqnarray}

\noindent we come to the conclusion that the $D$-dimensional potential energy can be computed through the simple expression

\begin{eqnarray}
E_D^{({\mathrm{scal}})}(r)=  \frac{\sigma_1 \sigma_2}{(2 \pi)^{D-1}} \int{d^{D-1}{\bf{k}}e^{i {\bf{k}} \cdot {\bf{r}}}
 D({\bf{k}})}.
\end{eqnarray}

It is worth noting   that Zee \cite{6}  utilized Eq. (1) to show  that in $D=4$  two static real scalar sources attract each other by virtue of their coupling to the scalar field.

On the other hand, the interparticle potential energy   for the interaction of two point-like static  electric charges in the framework of a $D$-dimensional electromagnetic model is given by \cite{1}

\begin{eqnarray}
E_D^{(\mathrm{electr})}(r)=  \frac{Q_1 Q_2}{(2 \pi)^{D-1}} \int{d^{D-1}{\bf{k}}e^{i {\bf{k}} \cdot {\bf{r}}}{\cal{P}}_\mathrm{0 0} ({\bf{k}})},
\end{eqnarray}

\noindent where ${\cal{P}}_{\mu \nu}(k)$ is the ``propagator" in momentum space obtained by neglecting all terms of the usual Feynman propagator in momentum space that are orthogonal to the external conserved currents and ${\cal{P}}_\mathrm{\mu \nu} ({\bf{k}}) \equiv {\cal{P}}_\mathrm{\mu \nu} (k)|_{k^0=0} $. We remark that in the deduction of (12) it was assumed that the external  current is conserved.
 
Let us then show that ${\cal{P}}_\mathrm{0 0} ({\bf{k}})= D({\bf k})$. To do that we recall that the Lagrangian associated with a $D$-dimensional electromagnetic theory can be written  as

\begin{equation}
{\cal L}^{(\mathrm{electr})}(x)= \frac{1}{2}A^\mu (x)O_{\mu \nu}(x)A^\nu(x).
\end{equation}

\noindent Here, $O_{\mu \nu}(x) \equiv a(x)\theta_{\mu \nu}(x) + b(x) \omega_{\mu \nu}(x)$ is the wave operator and $ \theta_{\mu \nu} \equiv \eta_{\mu \nu} - \frac{\partial_\mu \partial_\nu}{\Box}$ and $\omega_{\mu \nu} \equiv \frac{\partial_\mu \partial_\nu}{\Box}$ are the usual   vectorial projector operators. Accordingly, the corresponding propagator is given  by

\begin{equation}
O_{\mu \nu}^{-1}(x)= \frac{1}{a}\theta_{\mu \nu} + \frac{1}{b}\omega_{\mu \nu},
\end{equation}
 
\noindent which in momentum space can be written as

\begin{equation}
O_{\mu \nu}^{-1}(k)= \frac{1}{a(k)}\theta_{\mu \nu}(k) + \frac{1}{b(k)}\omega_{\mu \nu}(k),
\end{equation}
 \noindent where $\theta_{\mu \nu}(k) \equiv \eta_{\mu \nu} - \frac{k_\mu k_\nu}{k^2}$ and $\omega_{\mu \nu}(k) \equiv \frac{k_\mu k_\nu}{k^2}$.

Thus, ${\cal{P}}_{\mu \nu}({\bf k})= \frac{1}{a({\bf k})} \eta _{\mu \nu}$ and, as a consequence,

\begin{equation}
{\cal{P}}_{00}({\bf k}) = \frac{1}{a({\bf k})}.
\end{equation}

We formulate in the following a kind of correspondence principle in order to connect the electromagnetic and scalar fields. Technically, this link can be  achieved via the correspondence below 

$$A^\mu ({\mathrm {or}} \; \; A_\mu) \longrightarrow \phi,$$

$$\partial_\nu A^\mu \partial_\mu A^\nu  \longrightarrow  0,$$ 

$$J^\mu ({\mathrm {or}} \;\; J_\mu)\longrightarrow J.$$

\noindent We call attention to the fact that in the second  expression listed above, $\partial_\nu A^\mu \partial_\mu A^\nu$ stands for  all  the expressions that can be obtained from it via integration by parts. We also remark that as a straightforward consequence of the mentioned correspondence principle
 $ -\frac{1}{4}F_{\mu \nu}^2 \longrightarrow -\frac{1}{2} \partial_\mu  \phi \partial^\mu \phi$, where $F_{\mu \nu} \equiv \partial_\mu A_\nu - \partial_\nu A_\mu$ is the field strength.

Now, if we take  the  above mentioned correspondence principle into account, we promptly obtain from (13)

\begin{equation}
{\cal L}^{(\mathrm{scal})}(x)=\frac{1}{2}\phi(x)a(x)\phi(x).
\end{equation} 

Therefore, the scalar propagator in momentum space reads

\begin{equation}
D(k)= \frac{1}{a(k)}.
\end{equation}
 
\noindent As a result, $D({\bf k})= \frac{1}{a({\bf k})}$, implying that ${\cal{P}}_{00}({\bf k})=D({\bf k}) $.

Accordingly, the steps to be followed  to arrive at the $D$-dimensional interpaticle potential energy for an electromagnetic model  from the related scalar system, are:

\begin{itemize}
\item[A.] Use the correspondence principle to find the scalar Lagrangian from the original electromagnetic one.
\item[B.] Compute $D({ \bf k})$.
\item[C.] Calculate  $\int{d^{D-1} {\bf k} e^{i {\bf k}\cdot {\bf r}} D({\bf k})}$.
\item[D.] Obtain the $D$-dimensional interparticle potential energy $(E_D^{(\mathrm{electr})}(r))$ for the interaction of  two static  electric charges $(Q_1 \; {\mathrm{and}} \; Q_2)$ via the expression $$ \frac{Q_1 Q_2}{(2\pi)^{D-1}}\int{d^{D-1} {\bf k} e^{i {\bf k}\cdot {\bf r}} D({\bf k})}.$$
\end{itemize}

A comparison between the  scalar Lagrangian obtained via the correspondence  principle   and 
 that  related to the standard scalar field, clearly shows the they  differ by an overall minus sign (the correspondence principle changes the sign of the usual scalar Lagrangian); as a consequence, the standard scalar potential energy  between two static real scalar charges ($\sigma_1 \;{\mathrm{and}} \; \sigma_2$) must be computed  through the expression

\begin{eqnarray}
E_D^{({\mathrm{standscal}})}(r)=  -\frac{\sigma_1 \sigma_2}{(2 \pi)^{D-1}} \int{d^{D-1}{\bf{k}}e^{i {\bf{k}} \cdot {\bf{r}}}
 D({\bf{k}})}.
\end{eqnarray}

\section{$D$-dimensional electromagnetic  potential energy for a general model containing  higher derivatives from the corresponding scalar one}

To text the efficacy and simplicity of the method developed in the last section we shall find, using the aforementioned prescription,  the $D$-dimensional potential energy for the general electromagnetic model defined by the Lagrangian

\begin{eqnarray}
{\cal L}^{(\mathrm{electr})}&=& -\frac{1}{4}F_{\mu \nu}^2 + \frac{1}{2}M^2 A_\mu^2 - \frac{1}{4m^2}F_{\mu \nu} \Box F^{\mu \nu}  \nonumber \\ &&- J^\mu A_\mu, 
\end{eqnarray}

\noindent where $M $ and $m $ are parameters with mass dimension. Note that if  $M \rightarrow 0$ we recover Lee-Wick electrodynamics which has been object of increasing research as time goes by \cite{9,10,11,12,13,14,15,16,17,18,19,20,21,22,23,24,25,26,27,28,29,30,31,32,33}.  Actually, the electromagnetic model we are considering is nothing but a natural extension of the Lee-Wick system.   Since the method we  have developed assumes that the external current is conserved, for completeness sake,   we shall   find beforehand the constraint on the field $A^\mu$  owed to the mentioned  conservation law.

From (20), we immediately obtain

\begin{eqnarray}
\Big[1 + \frac{\Box}{m^2} \Big] \partial_\mu F^{\mu \nu} + M^2A^\nu=J^\nu.
\end{eqnarray}

 On the  other  hand,   requiring   conservation of the external current we arrive at the conclusion that $\partial_\mu A^\mu=0$. Note that if we  assume that the external current concerning  Proca electrodynamics is conserved, we obtain a constraint on $A^\mu$ that coincides with the one  we have just found. 

Taking the aforementioned restriction into  account, we get from (21) a  generalized wave equation for the field $A^\mu$, i.e.,

\begin{eqnarray}
\Big( \Box + \frac{\Box \Box}{m^2} + M^2 \Big)A^\mu=0.
\end{eqnarray}

After this little digression, let us return to our main theme: the  computation of the  potential energy  for an electromagnetic model from the  corresponding scalar system.

The scalar Lagrangian corresponding  to (20) can be easily obtained  by means of  the correspondence principle. The result is the following 

\begin{eqnarray}
{\cal L}^{(\mathrm{scal})}&=&-\frac{1}{2} \partial_\mu \phi \partial^\mu \phi + \frac{1}{2}M^2 \phi^2 - \frac{1}{2m^2}\partial_\mu \phi \Box \partial^\mu \phi \nonumber \\ &&-J\phi.
\end{eqnarray}

\noindent We recall, as we have already commented, that if we multiply the  above  Lagrangian by $-1$,  we will obtain the Lagrangian  describing the usual scalar model.  Therefore, the Lagrangian found via the correspondence principle is a  convenient   mathematical tool built out with the only purpose of allowing the computation of the electromagnetic potential energy through the use of a fictitious  scalar field.

 The propagator related to the model we are discussing can in turn be written in momentum space as

\begin{eqnarray}
O^{-1}= \frac{m^2}{k^4 -k^2 m^2 + m^2 M^2},
\end{eqnarray}

\noindent which implies that this system is endowed with  two massive poles, i.e.,

\begin{equation} 
m_{+}^2 \equiv \frac{m^2}{2}\Bigg[1+ \sqrt{1- \frac{4M^2}{m^2}}\Bigg],
\end{equation}

\begin{equation}
 m_{-}^2 \equiv \frac{m^2}{2}\Bigg[1 - \sqrt{1- \frac{4M^2}{m^2}}\Bigg].
\end{equation}

%\begin{eqnarray}
%O^{-1} (k)=\frac{1}{\sqrt{1- \frac{4M^2}{m^2}}}\Bigg[ \frac{1}{k^2 - m_{+}^2} - \frac{1}{k^2 - m_{-}^2} \Bigg],
%\end{eqnarray}

  To avoid tachyons in the model we assume that 
\begin{equation}
0\leq \frac{4M^2}{m^2} \leq 1.
\end{equation}

%It follows from (24) that

%\begin{eqnarray}
%D({\bf k})= \frac{1}{\sqrt{1 - \frac{4M^2}{m^2}}} \Bigg[ \frac{1}{ {\bf k}^2+ m_{-}^2} - \frac{1}{ {\bf k}^2+ m_{+}^2}\Bigg].
%\end{eqnarray}

Three interesting models arise from the constraint (27):

\begin{itemize}
\item[A.]  $M=0$: Lee-Wick electrodynamics.
\item[B.] $4M^2=m^2$: A model  in which the propagator has a pole of order $2$  at  $k^2= \frac{m^2}{2}$.
\item[C.] $0< \frac{4M^2}{m^2} < 1$: A system containing two modes with different non vanishing masses $m_{+}$ and  $m_{-}$.
\end{itemize}

We discuss each one of them in the following. 

\subsection{$M=0$}

In this case the model describes the celebrated Lee-Wick electrodynamics \cite{9,10}. Since this system has been considered in detail in Ref. 1, the main results of the research are summarized below.

\begin{enumerate}
\item The potential energy for the interaction of two static pointlike electric charges $Q_1$ and $Q_2$,  is given for $D=2,4,5, ...$ by

\begin{eqnarray}
E_D^{(\mathrm{electr})}(r) = &&\frac{Q_1Q_2}{(2\pi)^{\frac{D-1}{2}}} \Big[ \frac{2^{\frac{D-5}{2}} \Gamma (\frac{D-3}{2})}{r^{D-3}} \nonumber \\&& - \Big(\frac{m}{r} \Big)^{\frac{D-3}{2}} K_{\frac{D-3}{2}}(mr) \Big],
\end{eqnarray}
\noindent where $K_\nu$ is the modified Bessel function of the second kind of the order $\nu$; whereas for $D=3$ 

\begin{eqnarray}
E_3^{(\mathrm{electr})}(r) =- \frac{Q_1 Q_2}{2 \pi}\Big[\ln{\frac{r}{r_0}} + K_0(mr) \Big],
\end{eqnarray}

\noindent where $\Gamma$ is the gamma function, and $r_0$ is an infrared regulator.

\item Both (28) and (29) agree  asymptotically with the   Coulomb potential energy at large distances.

 \item For  $D=3$ and $D=4$ the higher derivatives present in the model are able to tame the wild  Coulombian divergence  at the  origin, while for $D=2$, $E_2^{(\mathrm{electr})}(r)= -Q_1 Q_2 \big[ \frac{r}{2} + \frac{1}{2m e^{mr}} \big]$ has a regular behavior at the origin.  Unluckily, for $D>4$  these higher derivatives  are unable to control this singularity at ${ r=0}$ . 

\end{enumerate}
\subsection{$4M^2=m^2$}

 Now, the propagator (24) reduces to

\begin{eqnarray}
 O^{-1}= \frac{m^2}{\Big(k^2 - \frac{m^2}{2}\Big)^2},
\end{eqnarray}

\noindent and, consequently,

\begin{eqnarray}
 D(k)= \frac{m^2}{\Big(k^2 - \frac{m^2}{2}\Big)^2},
\end{eqnarray}
which implies that

\begin{eqnarray}
 D({\bf{k}})= \frac{m^2}{\Big({\bf{k}}^2 + \frac{m^2}{2}\Big)^2}.
\end{eqnarray}

Therefore,

\begin{eqnarray}
 E_D^{(\mathrm{electr})}(r)= \frac{Q_1 Q_2 m^2}{(2 \pi)^{D-1}} \int{d^{D-1}{\bf{k}} \frac{e^{i {\bf{k}} \cdot {\bf{r}}}}{({\bf{k}}^2 + \frac{m^2}{2})^2}}.
\end{eqnarray}

Appealing  to Appendix A, we promptly obtain

\begin{eqnarray}
 E_D^{(\mathrm{electr})}(r)= \frac{Q_1 Q_2 }{(2 \pi)^{\frac{D-1}{2}}} \frac{m^2}{r^{\frac{D-3}{2}}}  {\int_0^\infty { \frac{x^{\frac{D-1}{2}}}{(x^2 + \frac{m^2}{2})^2} J_{\frac{D -3}{2}}(xr)dx}}, \nonumber
\end{eqnarray}
\noindent where $J_\nu$ is the Bessel function of the first kind. 

On the other hand, taking into account that

\begin{eqnarray}
\int_0^\infty \frac{J_\nu(bx) x^{\nu + 1}}{(x^2 + a^2)^{\mu + 1}} dx =\frac{a^{\nu -\mu} b^\mu}{2^\mu \Gamma(\mu +1)}K_{\nu -\mu}(ab),  
\end{eqnarray}
\noindent  where $-1 < \nu < (2\mu +\frac{3}{2})$,  we arrive at the conclusion that

\begin{eqnarray}
 E_D^{(\mathrm{electr})}(r)= \frac{Q_1 Q_2 m^{\frac{D-1}{2}}}{2^{\frac{3(D-1)}{4}} \pi^{\frac{D-1}{2}} r^{\frac{D-5}{2}}}  K_{\frac{D-5}{2}}\Big(\frac{mr}{\sqrt{2}}\Big), 
\end{eqnarray}
\noindent where   $2 < D <10$.

Keeping in mind  that $ K_\nu (r) \sim \sqrt {\frac{\pi}{2}} \frac{e^{-r}}{\sqrt{r}} \Big( 1 +  {\cal{O} }\Big( \frac{1}{r}\Big) \Big)$ for $r\longrightarrow \infty$, it is straightforward to see that (35)  and the   Coulomb potential energy agree asymptotically.  

Let us then  study the  behavior of the potential energy computed above  \big(see (35)\big) at short distances ($r \rightarrow 0$). To accomplish this task, we must consider two different situations: 
 \begin{itemize}
\item[A.] $\nu$ is equal to an integer  plus one-half ($\nu = n  +\frac{1}{2}$), which implies that 

\begin{eqnarray}
K_{n+\frac{1}{2}} (x) = \sqrt{\frac{\pi}{2x}} e^{-x} \sum \limits_{k=0}^{n} \frac{(n+k)!}{k! (n-k)! (2x)^k};
\end{eqnarray}
\noindent accordingly,   $K_{\pm \frac{1}{2}}(x)= \sqrt{\frac{\pi}{2x}}e^{-x}$.

\item[B.] $\nu$ is an integer, in  which  case

\begin{eqnarray}
K_\nu (x) = && (-1)^{\nu -1}\ln\Big(\frac{x}{2}\Big)  \Big(\frac{x}{2}\Big)^\nu   \sum \limits_{k=0}^{\infty} \frac{(\frac{x}{2})^{2k}}{k! (k+\nu)!} + \nonumber \\ && \frac{1}{2} \Big(\frac{2}{x}\Big)^\nu \sum \limits_{k=0}^{\nu - 1} \frac{(-1)^k (\nu - k -1)! }{k!}\Big(\frac{x}{2}\Big)^{2k}  + \nonumber \\ && \frac{(-1)^\nu}{2}\Big(\frac{x}{2}\Big)^\nu \sum \limits_{k=0}^{\infty} \frac{\psi(k+1) + \psi(k + 
\nu +1)}{k! (k + \nu)!}  \nonumber \\ && \times \Big(\frac{x}{2}\Big)^{2k}, \nonumber
\end{eqnarray}
\noindent where $\psi(x ) \equiv \frac{d}{dx}{\mathrm{ln }} \Gamma(x)$ is the psi function. As a consequence, if $\nu=0$.

\begin{eqnarray}
 K_0(x) = {\mathrm{ln}} \frac{1}{x} + ... \;. \nonumber
\end{eqnarray}
\noindent Here the dots denote  terms  remaining finite at $x=0$.

\end{itemize}

 Accordingly, if $D=3,4$ the potential is finite at the origin, whereas if $D=5,6, ...,9$ it has a singularity at $r=0$.

In Fig. 1 it is shown the behavior of the potential energy for the $D=3,4$ --- the cases where there are no singularities at the origin ---  while in Fig. 2 is depicted   the potential energy for all  the remaining  values of $D$ (5,6, ....,9). The reason for drawing two graphs was  to emphasize the difference between the  non singular and singular situations. Note, that in the singular cases,  $E_D^{(\mathrm{electr})}$ approaches  the singularity more slowly as $D$ increases.

\begin{figure}[h!]
	\centering
		\includegraphics[scale=0.5]{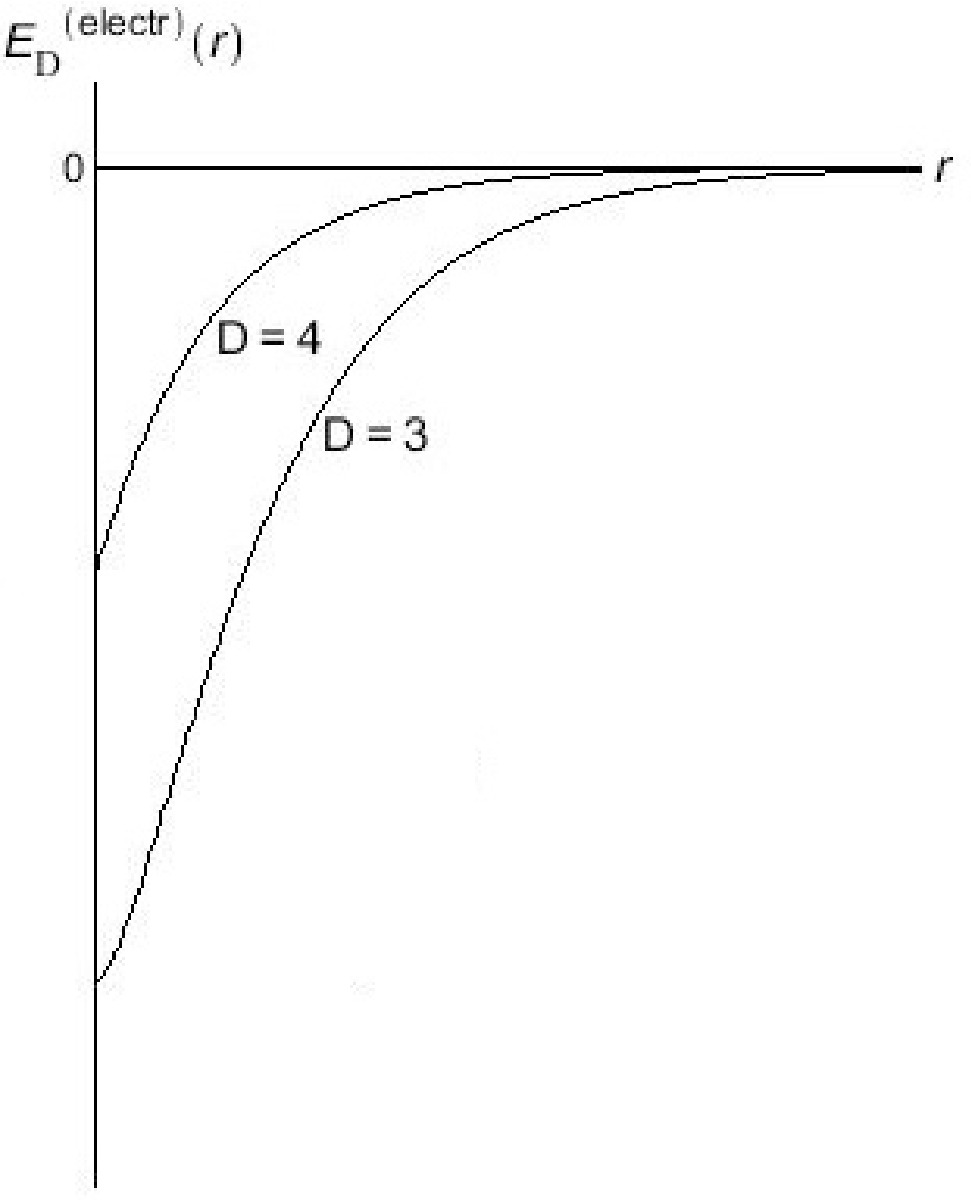}
	\caption{
	 Potential energy lacking singularity at the origin.}
	%\label{fig}
\end{figure}

\begin{figure}[h!]
	\centering
		\includegraphics[scale=0.5]{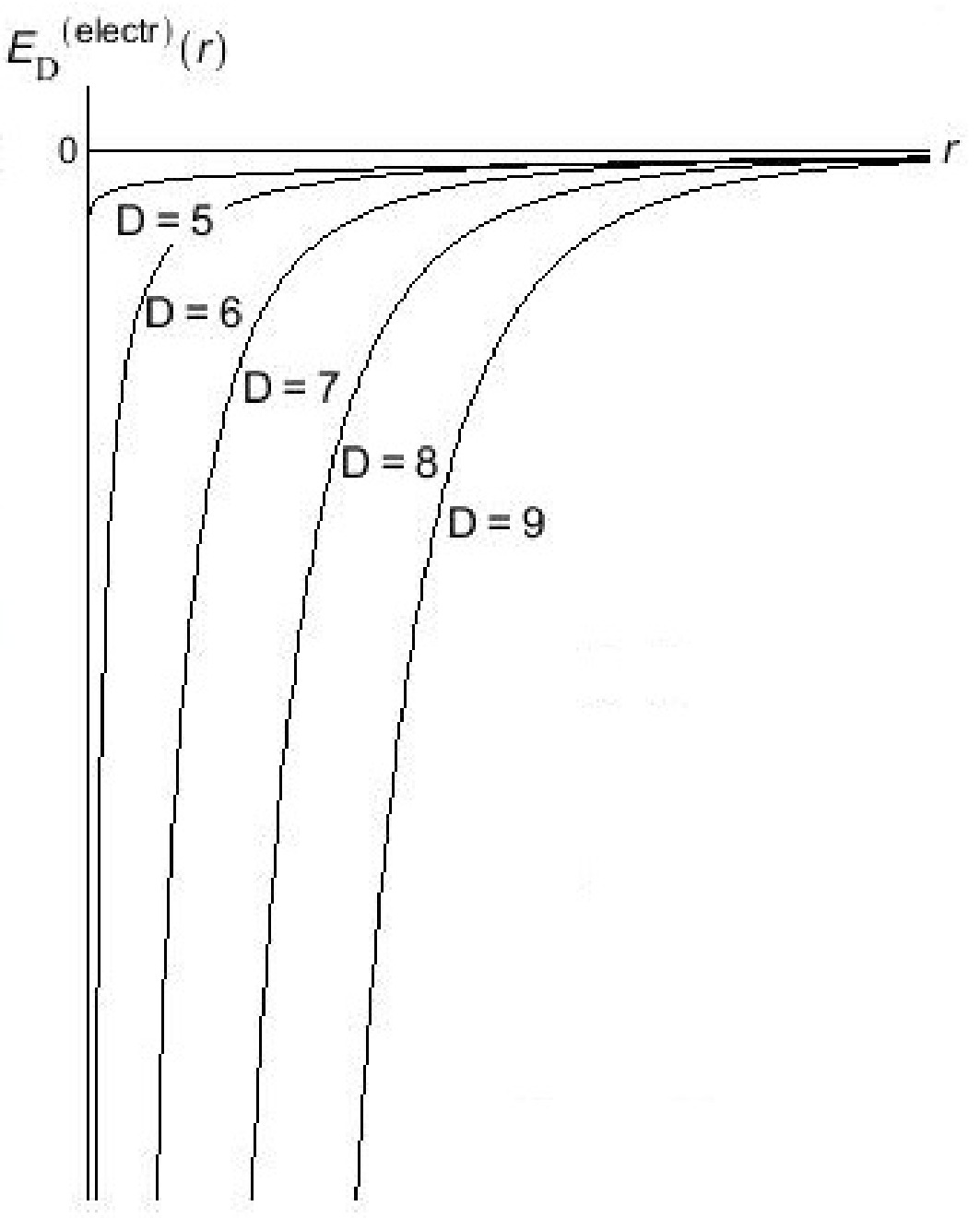}
	\caption{ Potential energy with  singularity at the origin.}
	%\label{fig}
\end{figure}

\subsection{$0< \frac{4M^2}{m^2} < 1$}
Here,

\begin{eqnarray}
E_D^{(\mathrm{electr})}(r)&&= \frac{Q_1 Q_2 }{(2 \pi)^{D-1}} \frac{1}{\sqrt{1 - \frac{4M^2}{m^2}}} \int d^{D-1}{\bf{k}} e^{i {\bf{k}} \cdot {\bf{r}}}\Bigg[ \frac{1}{{\bf{k}}^2 + m_{-}^2} \nonumber \\&&- \frac{1}{{\bf {k}}^2 + m_{+}^2} \Bigg]. 
\end{eqnarray}

Following the same steps as above, we promptly obtain

\begin{eqnarray}
E_D^{(\mathrm{electr})}(r)&&= \frac{Q_1 Q_2 }{(2 \pi)^{\frac{D-1}{2}}} \frac{1}{\sqrt{1 - \frac{4M^2}{m^2}}}
 \Bigg[\Big( \frac{m_{-}}{r}\Big)^{\frac{D-3}{2}} K_{\frac{D-3}{2}}(m_{-}r)   \nonumber \\ &&-  \Big( \frac{m_{+}}{r}\Big)^{\frac{D-3}{2}} K_{\frac{D-3}{2}}(m_{+}r) \Bigg], \;\;  2<D<6.
\end{eqnarray}

It is straightforward to see that (38) coincides asymptotically with the Coulombian  potential energy  at great distances. 

On the other hand, if $D=3,4$ the potential energy has no singularity at the origin, while if $D=5$ it diverges at this point.

Fig. 3 shows the behavior of $E_D^{(\mathrm{electr})}$ for $D=3,4,5$.

\begin{figure}[h!]
	\centering
		\includegraphics[scale=0.5]{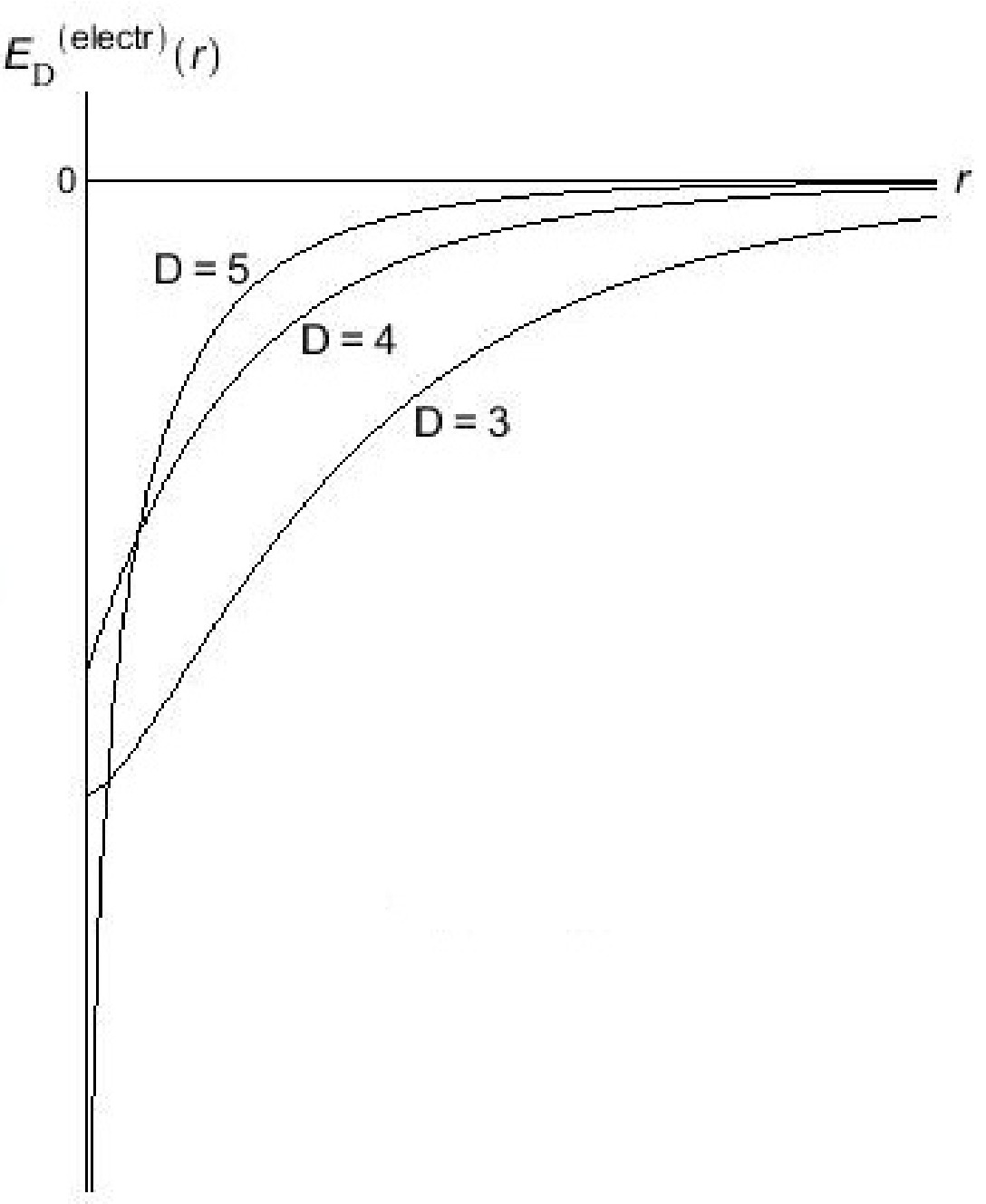}
	\caption{
	   Singular and  non singular potential energy at the origin.}
	\label{fig}
\end{figure}

\section{Ghosts in higher-derivative  theories}
The construction of regularized electrodynamics via the introduction of higher-order derivatives, was considered  by Bopp \cite{34}, Land\'{e} \cite{35,36,37}, and Podolsky \cite{38,39,40,41}, a long time ago. Currently, this method is employed in the regularization of both gauge \cite{42} and supersymmetric \cite{43} theories; higher-order derivatives are also a common ingredient in string theory \cite{44}.   

It is interesting to recall that to avoid divergences inherent in QED at short distances or, equivalently, at higher energies, we may introduce, for instance, a cutoff which renders the mass and charge of the electron finite. Indeed, consider in this spirit the Pauli-Villars regularization scheme used to obtain the electron self-energy. This prescription consists in cutoffing the integrals by assuming the existence of an auxiliary particle of heavy mass $m$. The propagator becomes modified by

\begin{eqnarray}
-\frac{\eta_{\mu \nu}}{k^2} \rightarrow \frac{\eta_{\mu \nu}}{k^2} \frac{m^2}{k^2 - m^2}= -\frac{\eta_{\mu \nu}}{k^2} + \frac{\eta_{\mu \nu}}{k^2 - m^2}. 
\end{eqnarray}
 \noindent The mass of the particle is related to a cutoff $l$, which tames the infinities of the theory, by $l=\frac{1}{m}$.  As the cutoff goes to zero, the mass of the auxiliary particle tends to infinity so that the unphysical fermion decouples from the system. 
 The above result clear shows  that an electromagnetic theory having a propagator equal to the that given by (39) must have a better behavior at short distances than the usual QED. On the other hand, it is easy to show that after adding a gauge-fixing Lagrangian, ${\cal {L}}_{\mathrm{gf}}= -\frac{1}{2\lambda} (\partial_\mu A^\mu)^2$, where  $\lambda$ is a gauge parameter, to the Lagrangian defining Lee-Wick theory (see the preceding section), i.e.,

\begin{eqnarray}
{\cal{L}}= -\frac{1}{4}F_{\mu \nu}^2 -\frac{1}{4m^2}F_{\mu \nu} \Box F^{\mu \nu},
\end{eqnarray}

 \noindent the resulting propagator,

\begin{eqnarray} 
D_{\mu \nu}(k)= \frac{m^2}{k^2(k^2 - m^2)}\Bigg[\eta_{\mu \nu} - \frac{k_\mu k_\nu}{k^2} \Bigg(1 + \lambda \Big(\frac{k^2- m^2}{m^2} \Big) \Bigg) \Bigg], \nonumber
\end{eqnarray}
\noindent coincides with (39) if the latter is sandwiched between conserved currents. Accordingly, the higher-order term of  Lee-Wick Lagrangian  modifies Maxwell  Lagrangian only at short distances, improving its ultraviolet behavior.

Unfortunately, there is a widespread prejudice against higher-order theories. In fact, many physicists have a strong, although unreasonable, bias towards these models. In general, two arguments are invoked  to discard these theories: 
\begin{itemize}
\item[A.]  Higher-order systems are  always plagued by ghosts. 
\item[B.] Ostrogradski's no-go theorem.
\end{itemize}
We do our best in the following to make these subjects more clear; in addition, we discuss  whether the four-dimensional version of the model found in Sec. III is infested by ghosts and analyze why the Coulomb singularity is lacking in this system.

\subsection{Demystifying the wrong idea that all higher-order models are haunted by ghosts}

Contrary to popular belief, not all higher-derivative systems are infected by ghosts. The following examples make clear that the  idea that higher-order models are always haunted by ghosts is fallacious. To avoid being prolix, we restrict our discussion to gravitational and electromagnetic higher-derivative models.

\subsubsection{ Higher-derivative gravity models}

In (2+1) dimensions,   the BHT model (``new massive gravity''), which is defined by the Lagrangian

\begin{equation}
{\cal{L}}=\sqrt{g} \Bigg[- \frac{2R}{\kappa^2} + \frac{2}{\kappa^2 m^2_2} \Bigg(R^2_{\mu \nu} - \frac{3}{8} R^2 \Bigg) \Bigg]    ,\nonumber
\end{equation}

\noindent where $\kappa^2=4\kappa_3$, with $\kappa_3$ being   Einstein's constant in (2 +1) dimensions, and $m_2$($>0$) is a mass parameter,  has no ghosts at the tree level \cite{45,46,47,48}. Interestingly enough, $R + R^2$ gravity in $(N +1)$ dimensions, i.e., the model defined by the Lagrangian ${\cal{L}} = \sqrt{(-1)^{N-2}g}\Big[\frac{2R}{\kappa^2} + \frac{\alpha}{2}R^2 \Big]$, where $\kappa^2= 4\kappa_{N + 1}$, with   $\kappa_{N + 1}$  being Einstein's constant in $(N + 1)$ dimensions, and $\alpha$ is a free parameter, is also  tree-level unitary \cite{49}.

 It is worth noting, that there exist models containing ghosts that there are harmless. An interesting example is found in the models studied by  Sotiriou and Faraoni: the so called $f(R)$ theories of gravity in (3 + 1)  dimensions. These authors analyzed these systems at the  the classical level and came to the conclusion that ``theories of the form $f(R, R^2, R^2_{\mu \nu})$, contains, in general, a massive spin-2 ghost field in addition to the usual massless graviton and the massive scalar'' \cite {50}. Nevertheless, at the linear level,  these theories are stable \cite{2}. The reason why they do not explode is because the ghost cannot accelerate  owing to energy conservation. Another way of seeing this is by analyzing  the free wave solutions.  Accordingly,  the  linear higher-derivative gravity model defined by linearizing  the Lagrangian

\begin{equation}
{\cal{L}}_1= \sqrt{-g} \Big[ \frac{2}{\kappa^2}R + \frac{\alpha}{2}R^2 + \frac{\beta}{2}R^2_{\mu \nu} \Big],
\end{equation}

\noindent where $\kappa^2= 4\kappa_4$, with $\kappa_4$ being Einstein's constant in (3+1) dimensions, and $\alpha$ and $\beta$ are free dimensionless coefficients, is not in disagreement with the result found by Sotiriou and Faraoni;  indeed, despite containing a massive spin-2 ghost, as asserted by these researchers, the alluded  ghost cannot cause trouble \cite{4}.  

Recently it was shown that at least in the  cases of specific  cosmological backgrounds, the unphysical  massive ghost  present in  the spectrum of  higher-derivative gravity in (3 + 1) dimensions is not growing up as a physical excitation and remains in the vacuum state until the initial frequency of the perturbation is close to  the Planck scale. Consequently, higher-derivative models of quantum gravity can be seen as very satisfactory effective theories of quantum gravity below the Planck cut-off \cite{3}. 

\subsubsection{Higher-order electromagnetic systems}
We begin by proving that Lee-Wick electrodynamics, although being infected by a ghost,  is tree-level unitary at familiar scales. To accomplish this task, we make use of a method pioneered  by Veltman \cite{51} that has been extensively used since it was conceived. The prescription consists in saturating the propagator with  external currents  and computing afterward the residues at all the poles of the alluded saturated propagator ($SP$). If the residues at the poles are positive or null, the model is tree-level unitary, but if at least one of the residues is negative, the system is nonunitary at the tree level.

The saturated propagator is momentum space is in turn given by
\begin{eqnarray}  
SP(k)&&=J_\mu(k)D^{\mu \nu}(k)J_\nu(k) \nonumber \\ &&= - \frac{J^\mu(k) J_\mu(k)}{k^2} + \frac{J^\mu(k) J_\mu(k)}{k^2 - m^2}.
 \end{eqnarray}
\noindent Here the external current is conserved.

Let us then suppose that $k^2 \ll m^2$. Consequently, $$ SP(k)= J_\mu J^\mu\Big[ - \frac{1}{k^2} \Big]+ {\cal{O}}\Big( \frac{k^2}{m^2}\Big).$$  Now, bearing in mind that 
 $$ \Big(J^\mu J_\mu \Big) \Big\vert_{k^2= 0} <0 \; ({\mathrm{see \; Ref. [49]}}),$$ we come to the conclusion that $$Res(SP)|_{k^2=0} >0.$$ Therefore, at the scale at  hand, Lee-Wick model is unitary at the tree level and, as a consequence,  the massive spin-1 ghost is  completely harmless.

We discuss now the tree-level unitarity of a  higher-derivative spin-1 model in (3 + 1) dimensions  built out by Dalmazi and Santos \cite{52}. The aforementioned system is of  the Maxwell-Proca type  and is defined by the Lagrangian

\begin{eqnarray}
{\cal{L}} = -\frac{1}{4}F_{\mu \nu}^2[\partial H] + \frac{m^2}{2} \Big(\partial^\nu H_{\mu \nu}\Big)^2 + H_{\mu \nu}J^{\mu \nu},
\end{eqnarray}

\noindent where $H_{\mu \nu}$ is a symmetric rank-two tensor, $F_{\mu \nu}[\partial H] = \partial_\mu \Big(\partial^\alpha H_{\nu\alpha}\Big) - \partial_{\nu }\Big(\partial^\alpha H_{\mu \alpha} \Big)$, and $J^{\mu \nu}= J^{\nu \mu}$ is the external current term which is not conserved. Note that if $J^{\mu \nu} =0 $, ${\cal{L}}$ is invariant under any local transformation preserving $\partial^\alpha H_{\alpha \beta}$; as a result,    ${\cal{L}}$ is invariant under the gauge transformation
$\delta_B H_{\mu \nu}= \partial^\sigma \partial^\rho B_{\mu \sigma \rho \nu}$, having the gauge parameters $B_{\mu \sigma \rho \nu}$ the same index symmetries of the Riemann tensor \cite{53}.

 Making the source term equal to zero and  adding the gauge- fixing term $\frac{\lambda}{2} G_{\mu \nu}^2 $ to the resulting  Lagrangian, where
\begin{eqnarray}
 G_{\mu \nu}(H)\equiv \Box H_{\mu \nu} - 2 \partial^\alpha \partial_{(\mu}H_{\nu) \alpha} + \eta_{\mu \nu}\partial^\alpha \partial^\beta H_{\alpha \beta}, \nonumber
\end{eqnarray}
 \noindent leads to

\begin{eqnarray}
{\cal{L}}= \frac{1}{2} H^{\mu \nu}O_{\mu \nu, \alpha \beta}H^{\alpha \beta},
\end{eqnarray}

\noindent where

\begin{eqnarray}
O(k)=&&\lambda k^4 P^{(2)} +\frac{k^2}{2 }(-k^2 + m^2)P^{(1)} + \lambda k^4 P^{(0-s)}  \nonumber \\ &&+ 
k^2((D-1)\lambda k^2 +m^2)P^{(0-w)} \nonumber \\ &&+ \sqrt{D-1} \lambda k^4(P^{(0-sw)}  + P^{(0-ws)}), \nonumber 
\end{eqnarray}
\noindent wherein $\{P^{(2)}, P^{(1)}, P^{(0-s)}, P^{(0-w)}, P^{(0-sw)},  P^{(0-ws)}\}$ is the set of the usual  $D$-dimensional Barnes-Rivers operators \cite{54}.

Accordingly, the propagator in momentum space reads

\begin{eqnarray}
O^{-1}(k)= &&i\Bigg[\frac{1}{\lambda k^4}P^{(2)} - \frac{2}{k^2(k^2 -m^2)}P^{(1)} + \frac{1}{k^2 m^2} P^{(0-w)} \nonumber \\&&+ \Bigg( \frac{1}{\lambda k^4} + \frac{D-1}{k^2 m^2} \Bigg) P^{(0-s)}  
- \frac{\sqrt{D-1}}{k^2 m^2} \Bigg( P^{(0-sw)} \nonumber \\ &&+ P^{(0-ws)} \Bigg)\Bigg]. 
\end{eqnarray}

Now, since the external current is not conserved, neither $k_\mu J^{\mu \nu}$ nor $k_\nu J^{\mu \nu}$ are null. Nonetheless, since the gauge symmetry is such that $\delta_B \partial^\nu H_{\mu \nu}=0$, the invariance of the source term $\delta_B \int{d^D x H_{\mu \nu }J^{\mu \nu}}=0$ requires $J_{\mu \nu}= \partial_\mu J_\nu + \partial_\nu J_\mu$, which in momentum space assumes the form $J_{\mu \nu}(k)= i(k_\mu J_\nu + k_\nu J_\mu)$. It follows then the saturated propagator can be written as

\begin{eqnarray}
SP(k)&=& J^{*}_{\mu \nu}(O^{-1})^{\mu \nu, \kappa\lambda}J_{\kappa \lambda} \nonumber \\ &=& iJ^{*}_{\mu \nu}\Bigg[ \frac{-2 P^{(1)}}{k^2(k^2- m^2)} + \frac{P^{(0-w)}}{m^2 k^2} \Bigg]^{\mu \nu, \kappa \lambda} J_{\kappa \lambda }\nonumber \\ &=& 4i\Bigg(  -\frac{J^{*}_\mu \theta^{\mu \nu} J_\nu}{k^2 - m^2} +\frac{ J^{*}_\mu w^{\mu \nu} J_\nu}{m^2}\Bigg).
\end{eqnarray}
\noindent The last line of (45) corresponds exactly to the two-point amplitude for the Maxwell-Proca model with non conserved external currents. Note that  the pole at $k^2=0$  cancels out and the imaginary part of the residue at $k^2=m^2$ is of course positive \cite{55}, which guarantees the tree-level unitarity of this higher-derivative Maxwell-Proca theory.

We remark that owing to the fact  that the external  current in momentum space related to the example above involves the presence of the imaginary unit $i$, the Veltman prescription  used in the first example had to be  reformulated as follows:

\begin{enumerate}
\item Add an $i$ to the propagator.

\item  Construct the saturated propagator  according to the following recipe: $$SP= {\mathrm {(external \;current)}}^{*} {\mathrm {(propagator)}}\; {\mathrm {(external \; current)}}.$$
\item Compute the residues at all the poles of   the imaginary part of $SP$; if these residues are positive or null, the model is tree-level unitary, but if at least one of the residues is negative, the system is non unitary at the tree level.
\end{enumerate}

 \subsection{Demystifying Ostrogradski's no-go theorem }
Let us then discuss the common misconception that {\it{singular}} higher-derivative models can be discarded by appealing to Ostrogradski's no-go theorem. For the sake of generality consider a  higher-derivative system in $D$ dimensions. According to popular belief, Ostrogradski's result implies that there exists a linear instability in the Hamiltonian associated with all higher-derivative systems. This is a completely untrue assertion. Indeed, Ostrogradski only treated non
singular models \cite{56}. Therefore, the only way of circumventing Ostrogradski's no-go theorem is by considering singular models, which is in accord with the conclusion reached by Woodard \cite{57}. An interesting example of this kind is the rigid relativistic particle studied by Plyushchay \cite{58}.

As is well known, all higher-derivative  gauge models are singular; as a result they can not be discarded {\it{a priori}} by  Ostrogradski's theorem since it does not apply to them. This does not mean, of course, that they are always ghost-free systems. In subsubsections 1 and 2, we exhibited some interesting higher -derivative gauge models that are tree-level unitary and obviously do not violate Ostrogradski's theorem.

\subsection{Counting ghosts}
We  investigate now whether the version in four dimensions of the model discussed in Sec. 3 is plagued by ghosts. This model, as it was shown in the aforementioned section, comprises three different cases  which were obtained  by imposing that  this system has no tachyons. We investigate these situations below.

\subsubsection{$M=0$}

Since in this case we are contemplating   Lee-Wick electrodynamics which is gauge invariant, Ostrogradski's theorem cannot be used to throw away this system. However, as we have already commented, this does not mean that it is lacking ghosts.
 A quick glace at (41) is enough to allows us to write
\begin{eqnarray}
Res(SP)|_{k^2=0} >0, \;\; Res(SP)|_{k^2= m^2} <0.
\end{eqnarray}

 \noindent Therefore, Lee-Wick electrodynamics is non unitary at the tree level; but if $k^2 \ll m^2$, this electrodynamics is unitary at this scale (see  Sec. IV). The latter situation is an example  of a model that despite having a ghost  is tree-level unitary. Actually, this ghost does not cause any trouble.

 We consider now the problem of the Coulombian singularity in Lee-Wick electrodynamics. Interestingly enough, it seems that there is a simple relation between the renormalizability of Lee-Wick theory and the absence of the  Coulombian singularity at the classical level:  the renormalizibibity implies that the potential energy is singularity free at the origin. This is  in accord with a conjecture that  appeared recently in the literature that says  that renormalizable higher-order models have  a classical potential energy lacking     singularities  at the origin  and are nonunitary \cite{59}.

 Indeed,  from (28)  we find that the modified Coulombian energy is

\begin{eqnarray}
E_{4}^{\mathrm{(electr)}}(r)=\frac{Q_1 Q_2}{4\pi}\Big[\frac{1}{r} - \frac{e^{-mr}}{r} \Big].
\end{eqnarray}

Expanding the exponential at the origin $r=0$ into powers series, it is  easy to check that the contribution of the higher-derivative term  to the Coulombian energy make it regular. The modified potential energy tends to the constant value

\begin{eqnarray}
E_{4}^{\mathrm{(electr)}}(r)=\frac{Q_1 Q_2}{4\pi}\Big[m + {\cal{O}}(r) \Big].
\end{eqnarray}
\noindent The singularity cancellation occurs because the zero order term of the Yukawa energy  produces the coefficient  -1 that cancels out the original Coulomb term.

Therefore, bearing in mind that the  Lee-Wick model is renormalizable, it agrees,  of course, 
with the cited conjecture.

\subsubsection{$4M^2=m^2$} 
The propagator concerning this system reads

\begin{eqnarray}
O_{\mu \nu}^{-1}(k) &&=i\Bigg[ \frac{m^2}{(k^2- \frac{m^2}{2})^2} \theta_{\mu \nu}
 + \frac{4}{m^2}\omega_{\mu \nu} \Bigg] \nonumber \\&&=  i\Bigg[ \frac{m^2}{(k_0^2- \omega^2)^2} \theta_{\mu \nu} + \frac{4}{m^2}\omega_{\mu \nu} \Bigg], 
\end{eqnarray}
\noindent where $ \omega^2\equiv {\bf{k}}^2 + \frac{m^2}{2}$.

Accordingly, the saturated propagator can be written as
 \begin{eqnarray}
SP(k_0, {\bf{k}})&&= i  J_\mu^*(k_0, {\bf{k}})
\frac{m^2}{(k_0^2 - \omega^2)^2} J^\mu(k_0, {\bf{k}}) \nonumber \\ &&= i  J_\mu^*(k_0, {\bf{k}})
\frac{m^2}{(k_0 - \omega)^2 (k_0 + \omega)^2 }J^\mu(k_0, {\bf{k}}), \nonumber
\end{eqnarray}

\noindent which implies that $SP(k_0, {\bf{k}})$ has two poles of  order two: one at $k_0=\omega$ and the other at $k_0 = - \omega$. As a result,

\begin{eqnarray}
Res(SP)|_{k_0=\omega}=\Bigg[\frac{d}{d k_0} \Bigg( \frac{i J_\mu^{*} J^\mu}{(k_0 + \omega)^2} \Bigg) \Bigg] _{k_0=\omega},
\end{eqnarray}

\begin{eqnarray}
Res(SP)|_{k_0=-\omega}=\Bigg[\frac{d}{d k_0} \Bigg( \frac{i J_\mu^{*} J^\mu}{(k_0 - \omega)^2} \Bigg) \Bigg] _{k_0=-\omega}.
\end{eqnarray}

In order to evaluate (51) and (52), we need beforehand the expression of the specific physical current concerning the model to be analyzed. Nevertheless, a way to get around this difficulty is to appeal to the conjecture mentioned in the preceding subsubsection. In fact, since this system is renormalizable, it is non unitary as far as this  conjecture is concerned, which also tells us that the classical
 potential energy is finite at the origin.  Let us prove this late assertion. From (35), we find

\begin{eqnarray}
E_4^{(\mathrm{electr}})(r)=\frac{Q_1 Q_2 m^{\frac{3}{2}}}{2^{\frac{9}{2}} \pi^{\frac{3}{2}}r^{-\frac{1}{2}}} K_{\frac{1}{2}}\Big({\frac{mr}{\sqrt{2}}}\Big).
\end{eqnarray}
\noindent Therefore, for $r \rightarrow 0$ the preceding result assumes the form 

\begin{eqnarray}
E_4^{(\mathrm{electr})}(r)=\frac{Q_1 Q_2 m}{4\sqrt{2} \pi} + {\cal{O}}(r),
\end{eqnarray}

\noindent which, of course, has no singularity at the origin.

\subsubsection{$0< \frac{4M^2}{m^2} <1$}
 Bearing in mind that the propagator for this system is given by

\begin{eqnarray}
(O^{-1})_{\mu \nu}= \Big[\frac{1}{k^2 - m_{+}^2} - \frac{1}{k^2 - m_{-}^2}\Big] \frac{\theta_{\mu \nu}}{\sqrt{1- \frac{4M^2}{m^2}}} + \frac{\omega_{\mu \nu}}{m^2}, \nonumber
\end{eqnarray}

\noindent we immediately find that

\begin{eqnarray}
Res(SP)|_{k^2= m_{+}^2} <0, \;\; Res(SP)|_{k^2= m_{-}^2} > 0.
\end{eqnarray}

Thus, the model is nonunitary; in addition, the potential energy is finite at the origin and equal to  $E_{4}^{({\mathrm{electr}})}(0)= \frac{Q_1 Q_2}{4 \pi \sqrt{1- \frac{4M^2}{m^2}}} (m_+ - m_-)$.

\subsubsection{Comment}
We have found that the  four-dimensional  electromagnetic  models described above are renormalizable,     nonunitary and  have a non-singular classical potential energy  at the origin, which agrees  with  a conjecture recently formulated \cite{59}. We remark that Stelle \cite{60} was the first to  hint at the possibility of existing a simple relation between the renormalizability of a  higher-derivative quantum theory and the absence of a singularity at the origin concerning the classical potential energy. Recently this subject was also  discussed in \cite{61,62}  Note, however that in \cite{59}, it is surmised that if a higher-order quantum theory is renormalizable, it has a classical potential energy finite at the origin and, besides, it is also nonunitary. 

\section{Final remarks}
We have developed  a simple prescription for computing the interaction potential energy between two  point like static charges in the framework of $D$-dimensional electromagnetic models from the corresponding scalar ones, via a correspondence principle that connects the electromagnetic and scalar fields. The key point of the method consists in finding the ``scalar propagator'' $D({\bf {k}})$, which is a quite trivial computation. The interparticle potential energy can then be easily calculated through the expression
\begin{eqnarray}
     E_D^{({\mathrm{elctr}})}  =  \frac{Q_1 Q_2}{(2\pi)^{D-1}}\int{d^{D-1} {\bf k} e^{i {\bf k}\cdot {\bf r}} D({\bf k})}.
\end{eqnarray}
 
This prescription allows also, as an added bonus, to compute the $D$-dimensional potential energy for the interaction of two scalar charges  through  the formula

\begin{eqnarray}
E_D^{({\mathrm{standscal}})}(r)=  -\frac{\sigma_1 \sigma_2}{(2 \pi)^{D-1}} \int{d^{D-1}{\bf{k}}e^{i {\bf{k}} \cdot {\bf{r}}}
 D({\bf{k}})}.
\end{eqnarray}

The method was  used  afterward to obtain the $D$-dimensional potential energy regarding a higher-derivative  electromagnetic  model, being  its behavior  discussed at  large as well small distances. It was  found that the four-dimensional systems that comprised  the mentioned model and which were obtained by requiring that they  were non tachyonic, are renormalizable, nonunitary, and have a potential energy that has no singularity at the origin. These results,  as we have already mentioned, are in complete accord with a conjecture that recently appeared in the literature \cite{59}.

It is interesting to note that if we have used the first order formalism to compute the  potential energy at the origin  alluded above,  we wold have  come to the conclusion that the  mentioned result is singularity free as well. Therefore, both prescriptions  lead
to the same conclusion.  Indeed, consider, for instance, Lee-Wick electrodynamics (see (20)). The field  theory with real vectorial fields $A_\mu$ and $Z_\mu$ with Lagrangian

\begin{eqnarray}
{\cal{L}}=&& \frac{1}{2}A_\mu \Box Z^\mu +\frac{1}{2}\partial_\mu A^\mu \partial_\nu Z^\nu - \frac{m^2}{8}A_\mu A^\mu  \nonumber \\  &&+ \frac{m^2}{4}A_\mu Z^\mu - \frac{m^2}{8}Z_\mu Z^\mu,
\end{eqnarray}

 \noindent is equivalent to the field theory with the Lagrangian (20). In fact, varying $Z_\mu$ gives
\begin{eqnarray}
Z_\mu= A_\mu + \frac{2}{m^2}\Box A_\mu - \frac{2}{m^2}\partial_\mu \partial_\nu A^\nu,
\end{eqnarray}
\noindent 
and the coupled second-order equations from (58) are fully equivalent to the fourth-order equations from (20). The system (58) now separates clearly into the Lagrangians for two fields, when we make the change of variables $A_\mu= B_\mu + C_\mu$ and $Z_\mu= B_\mu - C_\mu$. In terms of $B_\mu, \;C_\mu,\; B_{\mu \nu } \equiv \partial_\mu B_\nu - \partial_\nu B_\mu$, and $C_{\mu \nu} \equiv \partial_\mu C_\nu - \partial_\nu C_\mu$, the Lagrangian now becomes, 

\begin{eqnarray}
{\cal{L}}= -\frac{1}{4}B_{\mu \nu} B^{\mu \nu} + \frac{1}{4} C_{\mu \nu} C^{\mu \nu} - \frac{m^2}{2}C_\mu C^\mu,
\end{eqnarray}

\noindent which is nothing but the difference of the Maxwell Lagrangian for $B_\mu$ and the Proca Lagrangian for $C_\mu$. Accordingly, the potential energy for the the interaction of two static charges $Q_1$ and  $Q_2$  computed  using (60) coincides with (48), as a expected, being as a consequence   singularity free at the origin. We remark that Plyushchay  has employed  the first order formalism for dealing with mechanical systems with third derivatives \cite{58,63,64}. 

Could the prescription we have devised  be applied for the higher-derivative extension of the topologically massive electrodynamics build out by Deser and Jackiw \cite{65}? The answer is affirmative provided the external current is conserved. It is remarkable that theories with higher derivative Chern-Simons extensions can be related to non-commutative geometry \cite{66}.    

We then examined  the important  subject of ghosts in higher-derivative models in detail to clarify many prejudices against   these systems. In particular, we demystified  the wrong idea that all high
order models are infested by ghosts as well as the no-go theorem by Ostrogradski.

Last but non least, we would like to compare the potential energies obtained in subsection C (Counting ghosts) of section IV with the Coulombian energy. To do that we drew the graphs related to   these potential energies   together with the Coulombian potential. These pictures clearly shows that except in case A (Lee-Wick electrodynamics), there is a great difference between the remaining potentials  and the Coulomb one (see Fig 4). Why is this so? The answer, in a sense, is simple: Lee-Wick electrodynamics is the only linear  forth-order gauge-invariant generalization of Maxwell electrodynamics \cite{26}. As we have  already shown, the higher-order term of Lee-Wick Lagrangian modifies Maxwell Lagrangian only at short distances, improving its ultraviolet behavior.

%\begin{widetext}
 \begin{figure}[h!]
	\centering
		\includegraphics[scale=0.4]{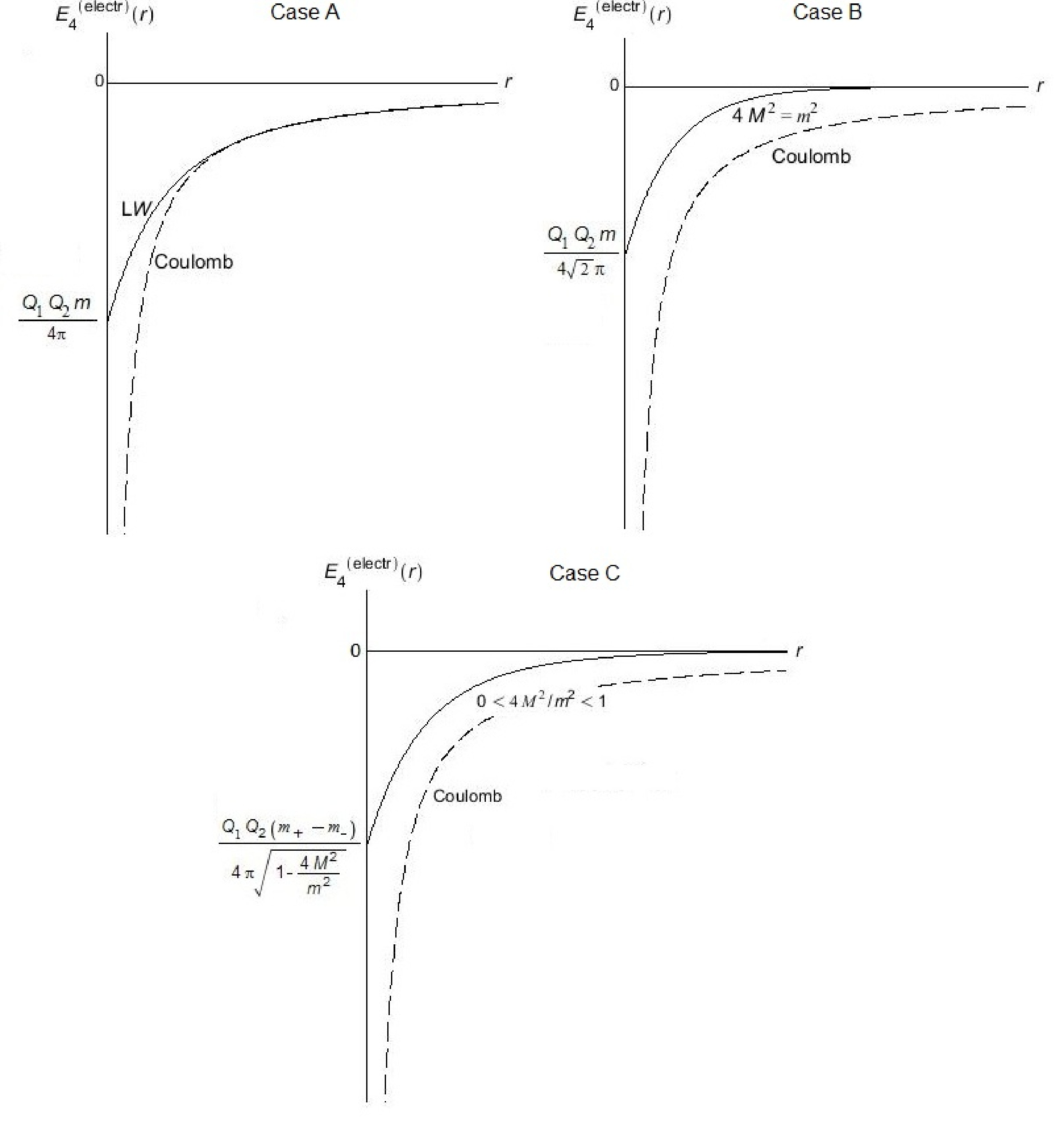}
	\caption{
	A comparison between the potential energy and the Coulomb one for the systems found in subsection C (Counting ghosts) of section IV.}
	\label{fig}
\end{figure}
%\end{widetext}

\begin{acknowledgments}
The authors acknowledge financial  support from CNPq and  FAPERJ.
\end{acknowledgments}

\appendix
\section{ SOLVING INTEGRALS OF THE TYPE    $\int{d^{D-1}{\bf{k}} e^{i{\bf{k} \cdot {\bf{r}}}} f(|{\bf{k}}|)}$}

In order to find the $D$- dimensional interparticle potential energy related to the models dealt with in this paper, we have to compute $E_D^{(\mathrm{electr})}(r)$ which can be generically  written as

\begin{eqnarray}
E_D^{(\mathrm{electr})}(r)= \frac{Q_1 Q_2}{(2 \pi)^{D -1}}\int{d^{D-1}{\bf{k}} e^{i{\bf{k} \cdot {\bf{r}}}} f(|{\bf{k}}|)}.
\end{eqnarray}

In this Appendix we show how to convert this expression --- which contains a  ($D-1$)-dimensional integral --- into a formula including only an uni-dimensional integral, which greatly facilitates the 
computational stage.

To begin with we introduce the variable ${\bf{x}} \equiv {\bf {k}}$ and represent the symbol $|{\bf{k}}|$ by   $x$. Using then the geometry depicted in Fig. 5 and   the  infinitesimal volume element in spherical coordinates $(
x, \theta_1, ..., \theta_{D-2})$, i.e.,

\begin{eqnarray}
d^{D-1} {\bf{k}} \equiv d^{D-1} {\bf{x}}= x^{D-2}dx \prod_{i=1}^{D-2} sin^{D-2-i} \theta_i d\theta_i,\nonumber
\end{eqnarray} 

\noindent $(A1)$ assumes the form

\begin{eqnarray}
E_D^{(\mathrm{electr})}(r)=&& \frac{Q_1 Q_2}{(2 \pi)^{D -1}} \Bigg[\int_0^{\infty} dx\; x^{D-2} f(x) \int_0^\pi  d\theta_1 e^{ixr cos\theta_1} \nonumber \\  &&\times   sin^{D-3} \theta_1 \Bigg]F,  \nonumber
\end{eqnarray}

\noindent where

\begin{eqnarray}
F \equiv&& \int_0^\pi d\theta_2 sin^{D - 4} \theta_2 \int_0^\pi d\theta_3 sin^{D-5} \theta_3   \nonumber \\ &&...  \int_0^\pi d\theta_{D-3} sin \theta_{D-3} \int_0^{2 \pi} d \theta_{D-2} \nonumber\\ = && \frac{2 \pi^{\frac{D-2}{2}}}{\Gamma\Big(\frac{D-2}{2}\Big)}. \nonumber
\end{eqnarray}

Now, bearing in mind that 

\begin{eqnarray}
\int_0^\pi e^{i \beta cosx} sin^{2 \nu} x dx= \sqrt{\pi} \Big(\frac{2}{\beta}\Big)^{ \nu} \Gamma \Big(\nu + \frac{1}{2} \Big) J_\nu (\beta), \; \nu > -\frac{1}{2}, \nonumber
\end{eqnarray}
 
\noindent we come to the conclusion that

\begin{eqnarray}
E_D^{(\mathrm{electr})}(r)= \frac{Q_1 Q_2}{(2 \pi)^{\frac{D -1}{2}} r^{\frac{D-3}{2}}} \int_0^\infty {x^{\frac{D-1}{2}} f(x) J_{\frac{D-3}{2}}(xr)dx}, \nonumber 
 \end{eqnarray}
\noindent where $D>2$. 
\begin{figure}[h!]
	\centering
		\includegraphics[scale=0.8]{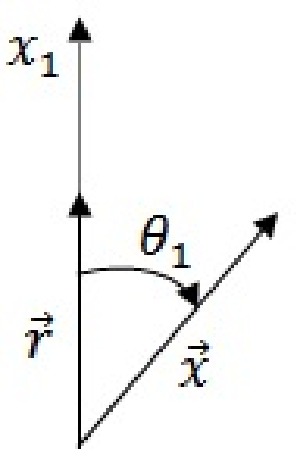}
	\caption{
	Geometry for the computation of the integral (A1).}
	\label{fig}
\end{figure}

\end{document}